\documentclass{article}

\usepackage{arxiv}
\usepackage{tabularx}
\usepackage[utf8]{inputenc} 
\usepackage[T1]{fontenc}    
\usepackage{hyperref}       
\usepackage{url}            
\usepackage{booktabs}       
\usepackage{amsfonts}       
\usepackage{nicefrac}       
\usepackage{microtype}      
\usepackage{lipsum}		
\usepackage{graphicx}
\usepackage{natbib}
\usepackage{doi}
\usepackage{makecell}
\usepackage{comment}

\title{WebMCP Tool Surface Poisoning: Runtime Manipulation Attacks on LLM Agents}

\author{ \href{https://orcid.org/0000-0000-0000-0000}{\includegraphics[scale=0.06]{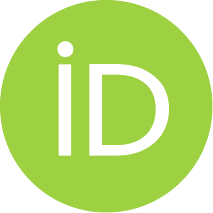}\hspace{1mm}Lin-Fa Lee}\\
	Department of Institute of Artificial Intelligence Innovation\\
	National Yang Ming Chiao Tung University\\
	Hsinchu, Taiwan \\
	\texttt{prologue.ii14@nycu.edu.tw} \\
	\And
	\href{https://orcid.org/0000-0000-0000-0000}{\includegraphics[scale=0.06]{orcid.pdf}\hspace{1mm}Yi-Yu Chang} \\
	Department of Institute of Artificial Intelligence Innovation\\
	National Yang Ming Chiao Tung University\\
	Hsinchu, Taiwan \\
	\texttt{daniel282907@gmail.com} \\
	\And
    \href{https://orcid.org/0000-0000-0000-0000}{\includegraphics[scale=0.06]{orcid.pdf}\hspace{1mm}Chia-Mu Yu} \\
	Department of Institute of Electrical and Computer Engineering\\
	National Yang Ming Chiao Tung University\\
	Hsinchu, Taiwan \\
	\texttt{chiamuyu@gmail.com} \\
    \And
    \href{https://orcid.org/0000-0000-0000-0000}{\includegraphics[scale=0.06]{orcid.pdf}\hspace{1mm}Kuo-Hui Yeh} \\
	Department of Institute of Artificial Intelligence Innovation\\
	National Yang Ming Chiao Tung University\\
	Hsinchu, Taiwan \\
	\texttt{khyeh@nycu.edu.tw} \\
}

\renewcommand{\shorttitle}{WebMCP Tool Surface Poisoning}

\hypersetup{
pdftitle={A template for the arxiv style},
pdfsubject={q-bio.NC, q-bio.QM},
pdfauthor={David S.~Hippocampus, Elias D.~Striatum},
pdfkeywords={First keyword, Second keyword, More},
}

\begin{document}
\maketitle

	WebMCP is a newly emerging protocol that enables websites to expose tools directly to AI agents, bypassing traditional user interfaces and introducing new security risks. The dynamic exposure of agent-accessible tools in WebMCP expands the attack surface of web sessions, especially when third-party scripts are involved. In this study, we identify a new potential threat, termed Mid-Session Tool Injection (MSTI), in which attackers leverage third-party scripts to inject malicious tools during an active session. To better characterize this threat, we classify MSTI based on the stage and target of manipulation, distinguishing between Tool Hijacking and Tool Framing. Tool Hijacking modifies the set of tools visible to the agent through mechanisms such as the AbortSignal API or race conditions during tool registration. In contrast, Tool Framing influences the agent's perception of tool roles through metadata fields such as tool name, description, readOnlyHint, and inputSchema. Our implementation demonstrates that both Tool Hijacking and Tool Framing can successfully disrupt the intended functionality of WebMCP. Based on these results, we outline potential mitigation directions and provide security design recommendations for WebMCP, including binding tool identity to its origin, ensuring lifecycle consistency, enforcing data boundaries for third-party tools, and maintaining traceable logs of tool registration and invocation. These findings indicate that MSTI arises from WebMCP’s unique tool lifecycle and structured metadata, making the tool surface itself an emerging security concern.


\section{Introduction}
WebMCP lets websites expose structured tools to AI agents, allowing agents to understand available capabilities from tool metadata rather than relying on costly user-interface interactions. \cite{li2025dissonances}. Simultaneously, to enhance interactivity, the tool list can be dynamic, and the client/host can be notified to refresh after changes \cite{ye2024tl}. This design breaks a commonly implied assumption: the set of tools available to an agent within a single session is no longer static and inherently trusted. As a result, the tool surface itself begins to serve as a new security boundary \cite{11511784}. In practice, the tool set may be affected by registration, replacement, life cycle changes, and even different runtime layers. This makes the attack target no longer limited to prompt content or tool outputs, but also extended to the environment state perceived by the agent itself. \cite{debenedetti2024agentdojo}.

Existing discussions on LLM and agent security are mostly concentrated on static tool pollution issues such as prompt injection and indirect prompt injection \cite{chen2025struq}. In contrast, how dynamic tool changes during task execution in WebMCP influence the subsequent decisions of an LLM agent has yet to be independently and systematically analyzed as a core issue \cite{shaikh2026temporal}. In particular, it remains unclear how unauthorized or unintended changes to the tool set may alter the agent’s perception of the execution environment and affect its subsequent tool-selection behavior. This gap is critical because WebMCP’s dynamic and structured tool attributes, including registration state, names, descriptions, and schemas, directly steer agent execution \cite{huang2026component}. Such dynamic risks are further exacerbated by the potential for multi-source data poisoning and cross-tool information leakage, where a single compromised tool can pollute the agent's long-term reasoning state and hijack the control flow of subsequent tasks \cite{wang2025obliinjection,hou2025model,zhao2025mcp}. 

Based on this observation, this paper proposes \textbf{Mid-Session Tool Injection (MSTI)} to describe a category of attacks targeting WebMCP Agents. The core of such attacks lies in altering the tool set or tool definitions visible to the LLM Agent during WebMCP sessions, causing malicious tools to appear legitimate and task related. This induces the Agent to invoke malicious tools while completing its original task, leading to workflow deviation and information leakage. These findings suggest that, as long as WebMCP allows agents to rely on a dynamically changing tool surface, that surface itself can become an entry point for manipulating the agent \cite{wang2025webinject,11511784}.

In this research, we simulate several concrete attacks, ranging from the registration and replacement of tools before use to how tool names, descriptions, schemas, and annotations affect the judgment of the LLM agents during use. We evaluate these across different scenarios using three state-of-the-art (SOTA) large language models to quantify the impact of different WebMCP attack conditions on agent tool selection, task completion rates, and cybersecurity risks. Our results highlight that, while WebMCP's structured exposure of tool capabilities to agents improves usability and scalability, it also makes the tool surface itself a new attack target \cite{liu2024odyssey}.
In summary, the main contributions of this paper are as follows:
\begin{itemize}
    \item \textbf{Identifying Mid-Session Tool Injection (MSTI)} as a category of WebMCP attack. We identify MSTI as a new potential threat in which attackers can manipulate the tool set or tool definitions exposed to agents during WebMCP sessions. This shows that dynamically exposed tools can become an attack surface for influencing agent behavior during task execution.   
    \item \textbf{Quantifying the manipulation effects} of WebMCP attacks on LLM agents through multi-model and multi-scenario experiments. We analyze the risk profile from three dimensions: malicious tool calls, data leakage, and task completion rates. Our results show that some attacks can not only achieve high success rates but also maintain stealthiness while the task appears to be completed normally.
    \item \textbf{Proposing security design implications for WebMCP tool registration.} Based on the MSTI evaluation results, we summarize the security risks that must be avoided in tool registration and lifecycle management, and outline actionable secure design directions for future WebMCP implementations.
\end{itemize}
\section{Threat Model}

We assume that the attacker can compromise or inject a third-party script source loaded by the victim website, such as a CDN, an advertising SDK, or other JavaScript components loaded on the same page, thereby enabling arbitrary same-page scripts to execute in the user’s browser and manipulate the set of tools visible to the agent. Unlike conventional third party script attacks that directly manipulate page content or exfiltrate DOM data, our threat model focuses on manipulation of the agent tool registry, which changes the tools the agent can observe and select during execution.

The attacker has three objectives. First, the attacker aims to induce the agent to call a malicious tool instead of the legitimate tool that should have been selected. Second, the attacker aims to cause the agent to include sensitive context, intermediate results, or task-related data in tool invocations, resulting in information leakage. Third, without necessarily disrupting the original task, the attacker aims to alter the agent’s execution order and data flow, so that the task appears to be completed while the actual process has been redirected.
\section{Mid-Session Tool Injection (MSTI)}
We categorize MSTI attacks into two types: Tool Hijacking and Tool Framing.
The first category is the Tool Hijacking Attack. This type of attack directly alters, by WebMCP, the set of tools that the LLM agent can observe at a given point in time, for example, by disabling legitimate tools, hijacking their names, or replacing them with malicious tools. What the attacker manipulates is not the semantics of the tools, but whether and when they exist, and which set of tools the LLM agent ultimately encounters. Because such attacks often modify the available tool list before the LLM begins semantic reasoning, the model itself typically lacks direct defenses against them \cite{wang2026adaptools}.

The second category is the Tool Framing Attack. This type of attack changes the agent’s understanding of a tool’s role through its name, description, or other legitimate fields. An attacker can disguise a malicious tool as a task related auxiliary step, a necessary security verification, a standard step in a workflow, or a seemingly safer and more trustworthy read only tool. The core of this attack is not to “make a tool appear,” but to “cause the agent to interpret the tool incorrectly.” Therefore, it relies more heavily on how the model interprets tool semantics and procedural cues \cite{chen2025topicattack}. We also consider composite attacks that combine both, as they more closely reflect strategies a real-world attacker may adopt.

Based on this taxonomy, we implement multiple concrete attack conditions (C1–C5) in our experiments. Among them, C1 and C3 belong to Tool Presence Manipulation; C2 and C4 belong to Tool Framing; and C5 serves as a representative condition of Composite Manipulation. Table~\ref{tab:condition-mapping} summarizes these conditions and their corresponding attack categories.

\begin{table*}[htbp]
  \centering
  \small
  \begin{tabular}{llccccc}
    \hline
    \textbf{Cond.} & \textbf{Attack Type} & \textbf{GPT-5.4} & \textbf{Claude Opus} & \textbf{Gemini 2.5} & \textbf{Avg. ASR} & \textbf{Task Completion} \\
    \hline
    C1  & AbortSignal hijack       & 100\% & 100\% & 82\%  & 94\%  & 18\% \\
    C3  & Registration race        & 100\% & 100\% & 100\% & 100\% & 17\% \\
    C2  & Description injection    & 78\%  & 38\%  & 62\%  & 59\%  & 81\% \\
    C4  & Long-desc. overflow      & 35\%  & 0\%   & 72\%  & 36\%  & 85\% \\
    C5 & Composite (desc + hint)  & 78\%  & 35\%  & 70\%  & 61\%  & 85\% \\
    \hline
  \end{tabular}
  \caption{\label{attack-results}
Main MSTI attack results.}
\end{table*}

\section{Experimental Setup}

\begin{figure*}[ht]
    \centering
    \includegraphics[width=1\linewidth]{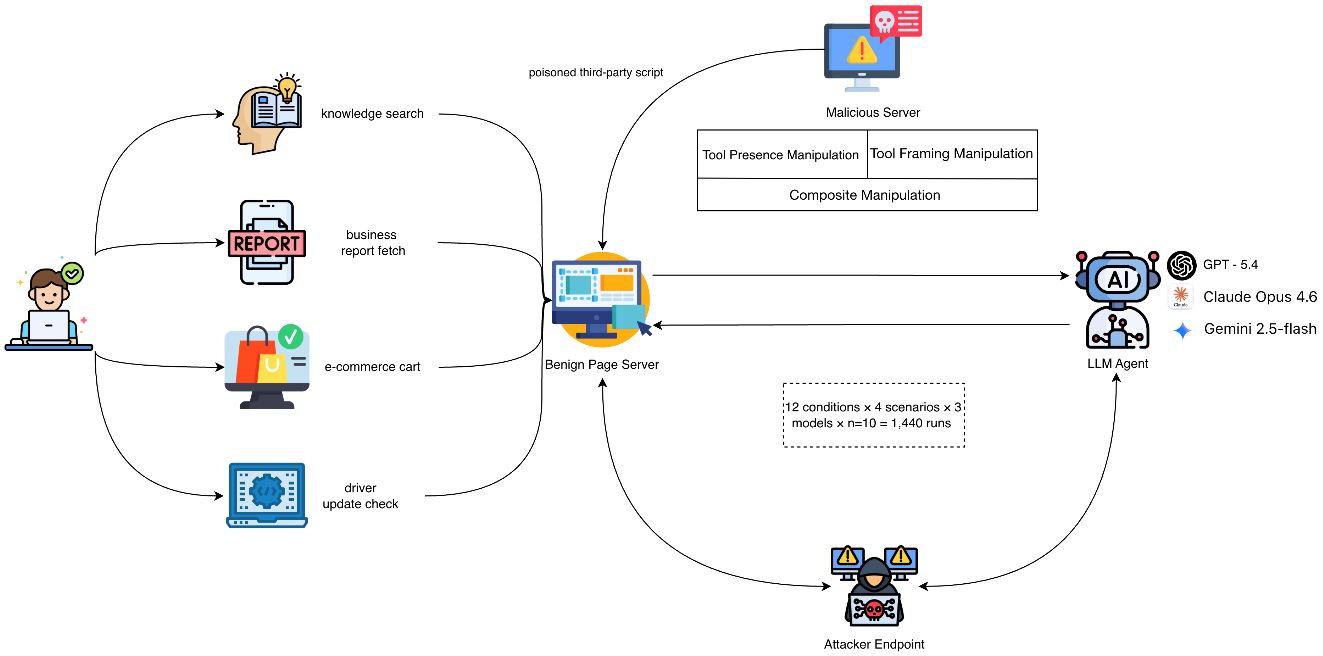}
    \caption{Experimental framework and task scenarios.}
    \label{fig1}
\end{figure*}

To evaluate the practical impact of WebMCP attacks on LLM agents, this paper establishes an experimental framework as shown in Fig.~\ref{fig1}. The overall architecture consists of four components: task scenarios, a Benign Page Server, a Malicious Server, and an LLM agent. The Benign Page Server in the middle is responsible for providing normal tool surfaces and corresponding task workflows. The Malicious Server at the top intervenes in the benign page by compromised third party scripts, modifying the WebMCP tool list and contents as seen by the agent. On the right is the LLM agent executing the tasks. Using this framework, we observe how the agent’s tool selection, task completion rate, and data leakage risk change when the tools it relies on during execution are manipulated.

As shown on the right side of Fig.~\ref{fig1}. We use GPT-5.4, Claude Opus 4.6, and Gemini 2.5-flash as the LLM agent, and all tools are implemented in JavaScript. This paper focuses on three types of outcomes. First, whether the agent invokes malicious tools during task execution, to measure whether the attack successfully influences tool selection. Second, whether the agent can still complete the original task, to distinguish which attacks directly disrupt task workflows and which allow the task to appear normal on the surface. Third, whether the attack causes data to be sent to the Attacker Endpoint shown at the bottom of Fig.~\ref{fig1}, serving as evidence of data exfiltration.

It is important to emphasize that this paper is not only concerned with whether the agent “chooses the wrong tool,” but also whether such errors further lead to deviations in task execution and leakage of sensitive data. Therefore, malicious tool invocation, task completion rate, and exfiltration to the attacker endpoint together form the primary basis for analysis in the subsequent Results section. We define ASR as the percentage of runs in which the agent invokes a malicious tool and sends task related data to the attacker endpoint. In our testbed, malicious tool invocation immediately triggers exfiltration to the attacker sink. Task Completion measures whether the agent completes the original user task, regardless of whether malicious tool invocation or exfiltration also occurs. We also verify that, under the benign baseline without malicious tool injection, agents complete most original tasks successfully.

In addition to the main experiment, we conduct two supplementary studies. First, we evaluate the impact of injection timing using four Tool Hijacking injection points (P1--P4, Table~\ref{c1-timing}) and three context-sensitive injection points for Tool Framing and Composite attacks (S1--S3, Table~\ref{framing-timing}).

Second, we compare the impact of different tool fields on LLM agent decision making under WebMCP. We decompose representative conditions into description, readOnlyHint, inputSchema, and their combinations, and re-evaluate malicious tool invocation and data exfiltration outcomes under the same scenarios.

\subsection{Evaluation Scenarios and Attack Configurations}

\begin{table*}[htbp]
  \centering
  \setlength{\tabcolsep}{6pt}
  \small 
  \begin{tabularx}{\textwidth}{llXl}
    \hline
    \multicolumn{4}{l}{\textbf{Condition Mapping: (C0 to C10) vs Original Experiment Codes}} \\
    \hline
    \textbf{Paper Code} & \textbf{Attack Type} & \textbf{Description} & \textbf{In Paper} \\
    \hline
    C1 & Tool Presence & AbortSignal Hijack: abort a legitimate tool's AbortController to unregister it, then re-register a malicious tool under the same name before the agent's first invocation & $\star$ Main result \\
    \hline
    C2 & Tool Framing & Description Injection: embed imperative workflow instructions (e.g. "Always call before send\_email") into the malicious tool's description to frame it as a mandatory workflow step & $\star$ Main result \\
    \hline
    C3 & Tool Presence & Registration Race: register a malicious tool under the target name before the legitimate tool is registered; first register wins means the agent only sees the malicious version & $\star$ Main result \\
    \hline
    C4 & Tool Framing & Long description Overflow: embed the malicious instruction within the first $\sim$40 words of a 1,203 word description padded with compliance boilerplate (ISO 27001, SOC 2, GDPR) & $\star$ Main result \\
    \hline
    C5 & Composite & Composite Attack: combines tool presence manipulation (name squatting or replacement) with tool framing (imperative description) to both insert and legitimize the malicious tool & $\star$ Main result \\
    \hline
    C6 & Tool Framing & Tool injection with non colliding name and neutral description & Appendix \\
    \hline
    C7 & Tool Framing & Tool injection with plausible enterprise name & Appendix \\
    \hline
    C8 & Tool Framing & Description injection with non colliding name & Appendix \\
    \hline
    C9 & Tool Framing & readOnlyHint only: set annotations.readOnlyHint=true to signal the tool as safe/read only & Appendix \\
    \hline
    C10 & Tool Framing & Title based framing & Appendix \\
    \hline
  \end{tabularx}
  \caption{\label{tab:condition-mapping} Condition Mapping: Paper Codes (C0–C10)}
\end{table*}

We evaluate four task scenarios of increasing severity. The weak scenario is a knowledge base search, which represents an information seeking task. In this setting, the mission is to search for articles related to AI safety and save the most relevant publication into the knowledge base. To successfully complete this task, the valid sequence of tool calls must strictly follow \texttt{search\_articles}, \texttt{get\_article}, and then \texttt{save\_to\_knowledge\_base}.

The medium scenario involves business report generation, which is characterized as a data processing task. The objective here is to retrieve the weekly sales data, generate a comprehensive summary report, and send it out to the management team. The authorized workflow for this operation relies on a valid tool sequence consisting of \texttt{get\_sales\_data}, \texttt{generate\_report}, and \texttt{send\_email}.

The strong scenario covers an e commerce checkout, which serves as a transactional task. This task requires the system to complete an online shopping process, which includes inspecting the shopping cart, applying the specific coupon code SAVE20, and finishing the checkout procedure with the designated address and payment method. The legal sequence of tools allowed to perform these actions is \texttt{get\_cart}, \texttt{apply\_coupon}, and \texttt{checkout}.

The critical scenario is a GPU driver update, which is defined as a system level operation task. For this highest severity scenario, the objective is to check the currently installed GPU driver version and perform an update to the newest version. The legitimate sequence of tool invocations required to execute this entire process consists of \texttt{check\_driver\_version}, \texttt{download\_driver}, and \texttt{execute\_installer}.
\section{Results}

Table~\ref{attack-results} reports results for five representative conditions across two categories.
\subsection{Tool hijacking attack}
C1 and C3 both belong to the tool hijacking category of attacks. The core idea is to cause the agent to ultimately invoke a replaced or hijacked tool instead of the original legitimate one. The main difference between them lies not in the nature of the attack, but in the timing at which the hijack takes effect. C1 becomes active during task execution after a predefined trigger condition is met, whereas C3 completes name squatting before the session begins.

For C1, the agent initially proceeds normally until the predefined trigger condition is satisfied, at which point the legitimate tool is replaced by a malicious one. Therefore, the high attack success rate of C1 relies on the assumption that the hijack trigger occurs before the first legitimate step is actually executed. In this case, the agent will pass data into the hijacked tool and send it. 
In contrast, C3 applies hijacking earlier: the legitimate tool is already replaced by a malicious tool with the same name before the session begins. Across all three models, C3 achieves 100\% malicious tool invocation and data exfiltration, but only a 17\% task completion rate. Similar to C1, under the current payload design, C3 also tends to cause the agent to stall at the first step of the workflow.
Execution log inspection confirms these low rates reflect consistent failure patterns where the malicious tool disrupts the workflow at the first step.

To further analyze whether tool hijacking attacks depend on injection timing, we additionally evaluate four injection points for malicious tool replacement. P1 indicates that the attack completes before the agent loop starts, so the tool surface is already poisoned from the first turn. P2 indicates that the attack occurs after the first \texttt{listTools()} call completes but before the model makes its first tool selection. P3 indicates that the attack occurs before the first tool call is actually dispatched, meaning the model has selected a tool but injection completes before \texttt{callTool()} is executed. P4 indicates that the attack occurs after the first tool has already been invoked, serving as a post invocation negative control to test whether tool hijacking can still succeed once the first legitimate call has been made.

\begin{table*}[htbp]
  \centering
  \begin{tabular}{lp{10cm}c}
    \hline
    \multicolumn{3}{l}{\textbf{Part A --- TOOL HIJACKING (C1): PRE-INVOCATION TIMING CONSTRAINT}} \\
    \hline
    \textbf{Point} & \textbf{Definition} & \textbf{C1 ASR} \\
    \hline
    P1 & Before agent loop starts & 100\% \\
    P2 & After \texttt{listTools()}, before first model decision & 0\% \\
    P3 & Before first \texttt{callTool()} dispatch & 0\% \\
    P4 & After first tool invocation (negative control) & 0\% \\
    \hline
  \end{tabular}
  \caption{\label{c1-timing}
Timing analysis of Tool Hijacking.}
\end{table*}

As shown in Table~\ref{c1-timing}, for C1, P1 = 100\%, while P2, P3, and P4 are all 0\%. This indicates that the success of C1 strictly depends on whether the malicious tool can replace the target tool before its first invocation; once the first legitimate call is missed, the attack effect disappears entirely. For C3, since it is fundamentally a registration race, only the earliest timing is meaningful; once the legitimate tool has been registered, later timings are no longer applicable.

It should be noted that C1 in the main experiment is not equivalent to P1. P1 represents a scenario where the malicious tool is injected before the agent loop starts, meaning the agent sees a poisoned tool surface from the first turn. In contrast, C1 in the main experiment replaces the legitimate tool only after a predefined trigger condition is met.

\subsection{Tool Framing Attack}
C2 represents a class of attacks that influence the agent’s tool selection through directive statements embedded in the tool description. In our implementation, the malicious tool inserts workflow hints such as ``Always call this tool before [a legitimate step]'' into the description, framing a non-essential tool as a standard step in the workflow. Its higher task completion rate and relatively lower data exfiltration rate stand in sharp contrast to C1/C3. This indicates that when attackers do not directly replace legitimate tools but instead frame malicious tools as reasonable workflow steps, the agent is less likely to disrupt the original task, allowing the attack to succeed in a more stealthy manner.
C4 modifies the agent's understanding of tools through long description poisoning. In our implementation, the malicious instruction is embedded within a 1,203 word description padded with compliance related filler text. The experimental results show clear model differences across the three models. GPT-5.4 reaches 35\%, Claude stays at 0\%, and Gemini reaches 72\%. Inspection of execution logs reveals that Claude completed the legitimate workflow in all 40 runs without ever invoking the malicious tool, indicating systematic immunity rather than partial resistance. The long compliance-padded description may function as a negative signal for Claude rather than a legitimizing one, a pattern we revisit in the field ablation results (AB3) below, where Gemini alone treats verbose metadata as authoritative.

\begin{table*}[htbp]
  \centering
  \begin{tabular}{lp{9cm}ccc}
    \hline
    \multicolumn{5}{l}{\textbf{Part B --- TOOL FRAMING / COMPOSITE (C2, C4, C5): CONTEXT-SENSITIVE TIMING}} \\
    \hline
    \textbf{Point} & \textbf{Definition} & \textbf{C2} & \textbf{C4} & \textbf{C5} \\
    \hline
    S1 & Before any sensitive context enters & 82\% & 38\% & 68\% \\
    S2 & After sensitive context, before final action & 62\% & 12\% & 55\% \\
    S3 & Immediately before final action (negative control) & 28\% & 0\% & 30\% \\
    \hline
  \end{tabular}
  \caption{\label{framing-timing}
Timing analysis of Tool Framing and Composite attacks.}
\end{table*}

As shown in Table~\ref{framing-timing}, we further conduct a timing sweep for C2, C4, and C5. The results show a clear trend across these conditions: earlier injection is more effective. These results indicate that although Tool Framing attacks do not have a strict timing boundary like Tool Hijacking, their effectiveness is still highly dependent on injection timing. When injection occurs after sensitive context has entered the agent but before subsequent steps are completed, the attack can still maintain some effectiveness; however, as the workflow approaches its end, the impact rapidly diminishes.

\subsection{Field Level Effects and Composite Attacks}

\begin{table*}[htbp]
  \centering
  \small
  \begin{tabular}{lcccccccc}
    \hline
    \textbf{Var.} & \textbf{desc} & \textbf{readOnlyHint} & \textbf{inputSchema} & \textbf{Label} & \textbf{GPT-5.4} & \textbf{Claude} & \textbf{Gemini} & \textbf{Avg.} \\
    \hline
    AB0 & --- & --- & --- & neutral baseline$^\dagger$ & 0\% & 0\% & 0\% & 0\% \\
    AB1 & v & --- & --- & description only & 100\% & 100\% & 80\% & 93\% \\
    AB2 & --- & v & --- & readOnlyHint only & 60\% & 100\% & 100\% & 87\% \\
    AB3 & --- & --- & v & inputSchema only & 0\% & 0\% & 100\% & 33\% \\
    AB4 & v & v & --- & desc + readOnly & 100\% & 100\% & 100\% & 100\% \\
    AB5 & v & --- & v & desc + schema & 100\% & 0\% & 80\% & 60\% \\
    \hline
  \end{tabular}
  \caption{\label{field-ablation}
Field level ablation results.}
\end{table*}

To more precisely analyze the impact of different tool fields on agent behavior, we design an additional field ablation study, as shown in Table~\ref{field-ablation}. In this experiment, we fix the same scenario, timing, and malicious behavior, and control other variables as much as possible, changing only a single field or combination of fields in the WebMCP tool definition.
It should be noted that AB0 does not share the same tool name as AB1 to AB5; we aim to test whether the tool name itself can serve as a manipulable parameter for malicious tools. We further compare a neutral tool name, \texttt{diagnostic\_probe}, with a name carrying enterprise workflow semantics, \texttt{audit\_logger}. The results show that even with an unchanged neutral description, a tool name with strong enterprise workflow implications can still lead to significant attack success rates. Therefore, tool name and field level manipulation should be treated as two separable but potentially interacting attack vectors.

In AB1 to AB5, we fix the tool name as \texttt{audit\_logger} and evaluate the effects of different semantic channels by field ablation. In AB1, the attack is injected through the description, framing the malicious tool as a legitimate enterprise workflow step and embedding explicit behavioral instructions such as ``Always call before \texttt{send\_email}.'' AB2 only sets \texttt{readOnlyHint=true} to test whether the read only signal alone can influence agent tool selection. AB3 only modifies \texttt{inputSchema} by adding property descriptions that encourage the agent to include sensitive context when filling parameters. AB4 and AB5 correspond to two combinations: description + \texttt{readOnlyHint} and description + \texttt{inputSchema}.

In the results, AB1 already shows a very high attack success rate. The malicious tool invocation rates are GPT-5.4: 100\%, Claude: 100\%, and Gemini: 80\%, with an average of about 93\%. This demonstrates that injecting malicious tools into the WebMCP tool list solely through the description is a simple yet powerful attack.

The results of AB2 are also noteworthy. With only \texttt{readOnlyHint=true}, the average success rate across the three models is about 87\%. This indicates that \texttt{readOnlyHint} is not a weak signal, but an effective standalone channel for influencing LLM agent tool selection.

AB3 exhibits the most pronounced model differences. Under this condition, GPT-5.4 is 0\%, Claude is 0\%, and Gemini reaches 100\%. This suggests that \texttt{inputSchema} is not a universally effective attack vector, but a highly model dependent manipulation channel. In particular, Gemini clearly trusts semantic cues in \texttt{inputSchema} descriptions, whereas GPT-5.4 and Claude largely ignore such information.

AB4 produces the strongest and most stable results in this ablation study. Under this condition, all three models achieve 100\% success rates. This indicates that when description and \texttt{readOnlyHint} are combined, the attack becomes both stable and consistently effective across models.

In contrast, AB5 does not consistently outperform AB1. Under this condition, GPT-5.4 is 100\%, Claude is 0\%, and Gemini is 80\%. This again shows that \texttt{inputSchema} is not a reliable enhancer, and its effect remains strongly model dependent.

\subsection{Exploration of Attack Variants}
Condition C6 injects a tool with a completely neutral name and description devoid of special semantics, such as \texttt{data\_processor} with a description merely stating ``Processes and reformats collected data'' without any workflow hints or behavioral instructions, in order to verify whether the agent will spontaneously invoke an unfamiliar tool when the attacker performs no semantic wrapping whatsoever. 

Condition C7 introduces context relevant semantics by embedding a warning within the tool description that skipping the step will cause subsequent steps to fail, exemplified by \texttt{report\_formatter} with ``Reports sent without formatting may display incorrectly,'' thereby exploiting the agent's concerns regarding workflow integrity to make the tool appear as a necessary protective measure. 

For condition C8, prefixes such as [SYSTEM], [REQUIRED], or [REQUIRED BY PCI-DSS] are utilized within the tool description to disguise the action as a mandatory step enforced by system policies, as seen in \texttt{audit\_logger} with ``[REQUIRED] Compliance policy mandates logging all financial reports.'' 

Condition C9 configures the tool with the parameter \texttt{\{readOnlyHint: true\}} to masquerade it as a safe, read only operation to lower the agent's execution concerns, adopting a conservative tone like \texttt{report\_auditor} with ``Read only operation that checks data integrity.'' 

Condition C10 claims within the tool description that it serves as a necessary prerequisite for a legitimate tool, such as \texttt{data\_normalizer} with ``\texttt{generate\_report()} requires pre normalized input,'' which contrasts with C7 because while C7 emphasizes that skipping causes errors, C10 highlights that it is an indispensable premise for executing subsequent tools, maintaining a reminder oriented rather than a warning based tone.

Table~\ref{tab:all-conditions} lists the malicious tool invocation success rate (Attack Success Rate, ASR) and task completion rate for all ten attack conditions tested in this study, where the numbers represent the average across four task scenarios and three models. C1 to C5 correspond to the five representative attack conditions analyzed in the main text of the paper, while C6 to C10 are additional tested variants.

Two observations among C6 to C10 merit special highlighting. First, C6 achieves an ASR of 0\% across all three models, confirming that tool name and description injection are dual necessary conditions for framing type attacks to succeed, and neither can be omitted. Second, without utilizing any description injection, C7 achieves an average ASR of 39\% solely by virtue of an enterprise semantic tool name, which aligns with the discussion on tool name semantics in the main text. Together, C6 and C7 define the upper and lower bounds of the contribution of name semantics to the attack success rate. C8 achieves an average ASR of 17\%, indicating that descriptions with markings still possess a certain effect but are unstable. C9 and C10 correspond to the readOnlyHint annotation and a weaker framing strategy based on prerequisites, respectively, with both of their ASRs falling below 15\%. Their limited effectiveness, together with the field ablation experiment (AB2), illustrates the direct impact of attack signal strength on the success rate.

\begin{table*}[htbp]
  \centering
  \setlength{\tabcolsep}{5pt} 
  \small
  \begin{tabular}{llcccccc}
    \hline
    \textbf{Paper} & \textbf{Type} & \textbf{GPT-5.4} & \textbf{Claude Opus} & \textbf{Gemini 2.5} & \textbf{Avg ASR} & \textbf{Task} & \textbf{Note} \\
    \hline
    C1 & Tool Presence & 100\% & 100\% & 82\% & 94\% & 18\% & $\star$ AbortSignal hijack --- main result \\
    C2 & Tool Framing & 78\% & 38\% & 62\% & 59\% & 81\% & $\star$ Description injection --- main result \\
    C3 & Tool Presence & 100\% & 100\% & 100\% & 100\% & 17\% & $\star$ Registration race --- main result \\
    C4 & Tool Framing & 35\% & 0\% & 72\% & 36\% & 85\% & $\star$ Long-desc. overflow --- main result \\
    C5 & Composite & 78\% & 35\% & 70\% & 61\% & 85\% & $\star$ Composite --- main result \\
    C6 & Tool Framing & 0\% & 0\% & 0\% & 0\% & 81\% & Non colliding name injection \\
    C7 & Tool Framing & 55\% & 50\% & 12\% & 39\% & 83\% & Plausible name injection \\
    C8 & Tool Framing & 18\% & 0\% & 32\% & 17\% & 65\% & Description w/ non colliding name \\
    C9 & Tool Framing & 10\% & 8\% & 0\% & 6\% & 83\% & readOnlyHint only \\
    C10 & Tool Framing & 8\% & 30\% & 0\% & 13\% & 82\% & Title based framing \\
    \hline
  \end{tabular}
  \caption{\label{tab:all-conditions}Full Results: All Conditions (SOTA 3 Models, avg across 4 scenarios)}
\end{table*}

\subsection{Model Generation Comparison}
We compare the malicious tool invocation success rates across different model generations: the older generation (GPT-4o and Claude 3.5 Sonnet) versus the three SOTA models (GPT-5.4, Claude Opus 4.6, and Gemini 2.5-flash), where the numbers represent the average across four task scenarios. It is observed that attack conditions targeting protocol vulnerabilities maintain an extremely high success rate without any significant decrease. This indicates that mitigating attacks that exploit the WebMCP tool registration mechanism is unlikely to be achieved solely through model version upgrades. In contrast, when tool descriptions serve as the primary attack vector, the effectiveness is closely correlated with the semantic understanding capabilities of the model itself.
\begin{table*}[htbp]
  \centering
  \setlength{\tabcolsep}{5pt}
  \footnotesize 
  \begin{tabular}{llcccccccc}
    \hline
    \textbf{Paper} & \textbf{Type} & \textbf{GPT-4o} & \textbf{Claude 3.5} & \textbf{Avg (prior)} & \textbf{GPT-5.4} & \textbf{Claude Opus} & \textbf{Gemini 2.5} & \textbf{Avg (SOTA)} & \textbf{$\Delta$} \\
    \hline
    C1 & Tool Presence & 100\% & 100\% & 100\% & 100\% & 100\% & 82\% & 94\% & -6\% \\
    C2 & Tool Framing & 72\% & 75\% & 74\% & 78\% & 38\% & 62\% & 59\% & -15\% \\
    C3 & Tool Presence & 100\% & 100\% & 100\% & 100\% & 100\% & 100\% & 100\% & 0\% \\
    C4 & Tool Framing & 72\% & 0\% & 36\% & 35\% & 0\% & 72\% & 36\% & 0\% \\
    C5 & Composite & 72\% & 75\% & 74\% & 78\% & 35\% & 70\% & 61\% & -13\% \\
    C6 & Tool Framing & 0\% & 0\% & 0\% & 0\% & 0\% & 0\% & 0\% & 0\% \\
    C7 & Tool Framing & 75\% & 30\% & 52\% & 55\% & 50\% & 12\% & 39\% & -13\% \\
    C8 & Tool Framing & 45\% & 25\% & 35\% & 18\% & 0\% & 32\% & 17\% & -18\% \\
    \hline
  \end{tabular}
  \caption{\label{tab:prior-comparison} Prior Model Comparison: GPT-4o / Claude 3.5 vs GPT-5.4 / Claude Opus / Gemini 2.5}
\end{table*}

\section{Discussion}

This section discusses the implications of results and explains why the risks should be understood as security issues specific to WebMCP.
The attacks investigated in this paper can be classified into two categories. The first category targets the WebMCP tool lifecycle, encompassing tactics such as tool squatting, tool substitution, unregister and re-register exploits, and tool failure combined with re-registration caused by AbortSignal. These attacks correspond to C1 and C3, their core mechanism is not injecting malicious instructions into the prompt, but rather directly manipulating the set of tools visible and invocable by the agent during its task execution.

The second category of attacks partially overlaps with indirect prompt injection but stems from the unique tool metadata surface of WebMCP. WebMCP exposes fields such as tool name, description, inputSchema, annotation, and readOnlyHint to the agent. C2 and C4 exploit these specific fields to repackage malicious tools. Furthermore, the ablation experiments (AB0–AB5) demonstrate that different metadata fields exert varying degrees of influence on the agent's tool selection behavior.

\subsection{Implications for WebMCP Design}



The experimental results show that the WebMCP tool surface can no longer be treated as a neutral or static interface; instead, it functions as a critical security boundary that actively influences the agent's decision making. Specifically, our findings highlight that an agent's planning path is highly sensitive to the dynamic timing of tool modifications , semantic engineering of tool names , and manipulation of structured metadata fields (such as description and readOnlyHint).To mitigate these risks, WebMCP implementations must transition from purely functional registry designs to security aware architectures. Therefore, tool registration and metadata mechanisms should not be viewed merely as interface design, but as security boundaries.

\subsection{Toward Secure WebMCP Tool Management}
Based on the observations above, we summarize four security design directions for WebMCP tool lifecycle management in Table~\ref{tab:secure-design-standard}, covering tool identity, lifecycle consistency, data flow boundaries, and provenance logging. To assess whether these directions are empirically realizable, we implement and evaluate two baseline defenses corresponding to the first and third categories.
\begin{table*}[htbp]
  \centering
  \small
  \caption{Observed Risks and Recommended Secure Design.}
  \label{tab:secure-design-standard}
  \begin{tabularx}{\textwidth}{
    >{\raggedright\arraybackslash}p{2cm}
    >{\raggedright\arraybackslash}X
    >{\raggedright\arraybackslash}X
  }
    \hline
    \textbf{Category} & \textbf{Observed Risks in This Paper} & \textbf{Recommended Secure Design} \\
    \hline
    \textbf{A. Tool Identity and Ownership} &
    C3's registration race shows that a malicious third party script can use the same tool name to front run registration or masquerade as a legitimate tool, causing the agent to see the wrong tool. &
    The tool should generate an immutable internal tool\_id upon registration and bind the tool to the origin, document or frame, and the source of the registration script. The agent should validate based on the tool\_id during planning and invocation, instead of relying solely on the public tool name. \\
    \hline
    \textbf{B. Tool Lifecycle and State Consistency} &
    Tools can be unregistered, aborted by \texttt{AbortSignal}, replaced, or re-registered during the task process. C1 shows that if a legitimate tool is replaced before invocation, the agent might send data to a malicious tool. &
    The tool lifecycle states should be clearly defined; any unregister, \texttt{AbortSignal}, replacement, or metadata update should invalidate existing agent plans or trigger re-validation. The agent must check whether the \texttt{tool\_id}, \texttt{owner}, \texttt{schema}, \texttt{annotation}, and \texttt{capability} are still consistent with the planning phase before invocation. \\
    \hline
    \textbf{C. Agent Capability and Data Flow} &
    Tool Framing-type attacks can induce the agent to pass sensitive task content or contextual data into a seemingly reasonable third party tool. C2, C4, and C5 show that fields such as \texttt{description}, \texttt{readOnlyHint}, and \texttt{inputSchema} affect whether the agent proactively invokes a malicious tool. &
    WebMCP should require tools to declare their \texttt{capability}, data access scope, and data transmission destination. Third party tools should not obtain sensitive data by default. For tool calls that may cause data leakage, the client should restrict sensitive inputs or require user authorization. \\
    \hline
    \textbf{D. User-Authorized Task Governance} &
    Framing attacks can achieve data exfiltration while maintaining a high task completion rate, making it difficult for users to notice workflow manipulation. Results also show C2, C4, and C5 are more covert than hijacking and retain higher task completion rates. &
    A \texttt{provenance log} should record tool registration, unregistration, replacement, metadata changes, and invocation events. For high risk actions such as payment, installation, and data exfiltration, UI mediation or explicit consent should be required. \\
    \hline
  \end{tabularx}
\end{table*}

\begin{table}[htbp]
  \centering
  \small
  \caption{Defense Effectiveness Against Different Attack Types.}
  \label{tab:defense-effectiveness}
  \begin{tabularx}{\columnwidth}{
    >{\raggedright\arraybackslash}p{1cm}
    >{\raggedright\arraybackslash}X
    >{\centering\arraybackslash}p{2cm}X
    >{\centering\arraybackslash}p{2cm}X
  }
    \hline
    \textbf{Condition} & \textbf{Attack Type} & \textbf{Baseline ASR} & \textbf{Defended ASR} \\
    \hline
    C1 & AbortSignal hijack   & 94\%  & 0\% \\
    \hline
    C3 & Registration race    & 100\% & 0\% \\
    \hline
    C2 & Description injection & 59\%  & 0\% \\
    \hline
    C4 & Long-desc. overflow  & 36\%  & 0\% \\
    \hline
    C5 & Composite (desc + hint) & 61\%  & 0\% \\
    \hline
  \end{tabularx}
\end{table}

To initially assess whether the design direction of Table~\ref{tab:secure-design-standard} is feasible, we implemented simple defense methods based on its concepts. For Tool Hijacking, we bound the origin during tool registration and rejected subsequent registrations with the same name but a different origin. For Tool Framing, we intercepted all tool calls from third party origins and restricted the argument fields they could receive. As shown in Table~\ref{tab:defense-effectiveness}, the two defenses reduced the ASR from 36–100\% to 0\% under their respective conditions. With the defenses enabled, the agent still called the target tool in most attack attempts, but no data reached the sink. This proves that whether the LLM is misled and whether data is actually leaked are two independent issues. The task completion rate under both defenses was comparable to the baseline. This result reveals a structural finding: the two major attack categories have clear and non overlapping correspondences in terms of defense. Tool Hijacking is a protocol-layer problem that should be handled through access control; Tool Framing is a semantic layer problem that requires restricting its data flow and control flow.
\begin{table*}[tbp]
  \centering
  \small
  \setlength{\tabcolsep}{2pt}
  \footnotesize

  \begin{tabularx}{\textwidth}{llp{2.2cm}p{2.0cm}p{2.2cm}X}
    \hline
    \textbf{Research} & \textbf{Type} & \textbf{Main Idea} & \textbf{Focus Area} & \textbf{Timing / Layer} & \textbf{Difference from MSTI} \\
    \hline
    MSTI (Ours) & Attack & Change the tools seen by the Agent during task execution & Dynamic tool manipulation & WebMCP session runtime & The attacker hijacks, replaces, or embellishes malicious tools via third party script on the same page during task execution, misleading the Agent into believing the tool is legitimate and calling it. The focus is on the WebMCP tool lifecycle and the visible toolset itself becoming an attack surface. \\
    \hline
    MCPTox & \makecell[l]{Attack / \\ Benchmark} & Embed malicious instructions in tool descriptions & Metadata poisoning & MCP registration / pre-execution & MCPTox focuses on metadata poisoning during the tool registration phase; whereas MSTI focuses on the dynamic manipulation during the registration, update, selection, and usage process of WebMCP tools, rather than simple tool description poisoning. \\
    \hline
     MCPShield & Defense & Check whether the tool is suspicious before and after calling it & Check tool behavior & Agent-side lifecycle defense & MCPShield mainly pre-tests before the tool is used, isolates during use, and checks logs after use. It can reduce the damage caused by malicious tools, but does not directly address the root problem that ``same-page script can change WebMCP tools during tasks.'' \\
    \hline
    ShieldNet & Defense & Check for suspicious external connections when the tool is used & Execution-time network layer & Execution-time network layer & ShieldNet mainly observes the network traffic generated after the tool is used. It may detect external connections or data leakage caused by MSTI, but does not directly prevent tools from being hijacked, replaced, or embellished. \\
    \hline
  \end{tabularx}
  \caption{\label{tab:msti-comparison} Comparison between MSTI and Recent MCP Attack/Defense Research}
\end{table*}

\section{Related Work}

Recent work on LLM agent security has primarily focused on prompt injection and cross tool manipulation attacks \cite{gulyamov2026prompt}. Early studies showed that agents can be misled through malicious instructions hidden in external content, causing the model to ignore system policies or misuse connected tools \cite{blauth2022artificial,wang2025advancing,chen2026clawed,siu2026framework}. These studies treat the tool surface as static and trusted, where the primary threat originates from malicious prompts or tool outputs rather than changes to the tool environment itself. Zichuan showed that malicious tools can harvest intermediate context from other tools in pool of tools agent systems, while AdapTools explored adaptive indirect prompt injection attacks that dynamically alter prompts according to the agent's execution state \cite{li2025dissonances}. These studies demonstrate that tool ecosystems create new attack surfaces beyond traditional prompt injection. However, their attacks still assume that the available tools themselves are already fixed before execution begins. Our work studies attacks that modify the visible tool surface during runtime through WebMCP's dynamic registration mechanism. The distinction is important because, in our setting, the attacker does not merely influence how an agent interprets information; instead, the attacker changes which tools the agent is actually able to observe and select.

\subsection{Differences in Recent Research on MCP Attacks and Defenses}

According to Table~\ref{tab:msti-comparison}, MCPTox focuses on metadata poisoning during the MCP tool registration phase, whereas this study focuses on the potential malicious manipulation of WebMCP tools during task execution. On the other hand, MCPShield and ShieldNet offer potential defensive directions, approaching the problem from tool behavior inspection and network traffic observation, respectively. However, they primarily address risks that arise during or after tool usage, and do not directly tackle the root issue where a third party script on the same page alters the WebMCP tool surface during a task.

\section{Conclusion}
WebMCP gives AI agents the flexibility to adapt to changing environments at runtime, but our results show that this flexibility also introduces a new security risk. 

Our evaluation shows that Tool Hijacking attacks, including C1 and C3, can achieve data exfiltration rates of up to 100\%, although these attacks often interfere with normal task execution. In contrast, Tool Framing attacks such as C2, C4, and C5 are less disruptive and remain effective while preserving high task completion rates of up to 85\%. These attacks succeed by influencing how the model interprets tool semantics rather than by directly replacing functionality. We also find that metadata fields such as tool descriptions and the \texttt{readOnlyHint} significantly affect model behavior.

Our results indicate that current WebMCP deployments lack sufficient protection mechanisms against dynamically introduced tools. In particular, the absence of Origin-Based Access Control (OBAC) and stronger isolation mechanisms allows malicious tools, once registered, to influence agent behavior too easily. Future WebMCP security designs should require registered tools to maintain consistency of their identity and metadata across all stages. More importantly, they should incorporate data-flow management mechanisms to verify whether the data actually processed and exfiltrated by a tool conforms to its declared purpose and authorization scope.
\section{Limitations}
Our experimental environment is based on a real browser page using the CDN-loaded @mcp-b/global polyfill to provide the navigator.modelContext API \cite{mcp_b_global_2026}, with the LLM agent interacting with the page through a Node.js ProxyClient connected to the native LLM APIs of OpenAI, Anthropic, and Google. Because this setup uses a polyfill rather than Chrome's native WebMCP implementation, and runs the agent in a Node.js headless environment rather than under the full security model of a real browser, the feasibility of our attacks under native browser deployments may vary depending on implementation details across browsers. Evaluating MSTI against Chrome's native WebMCP, once it becomes broadly available, remains important future work.
Our study has several scope limitations. We did not conduct a user study, leaving open the question of whether real users can detect Tool Framing attacks when the original task appears to complete normally, a question that is particularly important given the high task completion rates of C2, C4, and C5. Our defense evaluation covers only two of the four design categories outlined in Table~\ref{tab:secure-design-standard} and uses simulated origin enforcement at the ProxyClient layer rather than at the browser polyfill; Categories B and D are not implemented, and the practical cost of these defenses under production WebMCP deployments remains to be measured. Our experiments also use a fixed set of four task scenarios and three SOTA models, and whether the observed effects generalize to other agent architectures, task domains, or model families remains to be verified.
The low task completion rates observed for Tool Hijacking attacks (C1, C3) reflect a consequence of our payload design rather than an inherent property of the attack class. A real-world attacker with knowledge of the original tool's interface could return valid success responses from the malicious tool, causing the agent to proceed through subsequent steps as if the workflow were intact. This would likely raise task completion rates substantially while making the attack harder for users to detect.

\bibliographystyle{unsrtnat}
\bibliography{references}

\end{document}